\documentclass[12pt]{article}
\usepackage{graphicx}
\begin{document}
\noindent
\vspace{2cm}
\begin{center}
\textbf{\LARGE Quantum thermal waves\\ in quantum corrals}

\vspace{2cm}
{\Large Janina Marciak-Kozlowska, Miroslaw Kozlowski}

\bigskip
Institute of Electron Technology, Al. Lotnik\'{o}w 32/46, 02-668 Warsaw, Poland

\end{center}
\vspace{2cm}

\begin{abstract}
In this paper the possibility of the generation of the thermal waves in 2D electron gas is investigated. In the frame of the quantum heat transport theory the 2D quantum hyperbolic heat transfer equation is formulated and numerically solved. The obtained solutions are the thermal waves in electron 2D gases. As an example the thermal waves in quantum corrals are described.

\end{abstract}

\newpage
\section{Introduction}
Recently has been a great interest in both ultrafast (femtosecond and attosecond) laser-induced kinetics and in nanoscale properties of matter. Particular attention has been attracted by phenomena that are simultaneously nanoscale and ultrafast~\cite{1}-\cite{9}. Fundamentally nanosize eliminates effects of electromagnetic retardation and thus facilitates coherent ultrafast kinetics. On the applied side, nanoscale design of optoelectronic devices is justified if their operating times are ultrashort to allow for ultrafast computing and transmission of information.

One of the key problems of ultrafast/nanoscale physics is ultrafast excitation of nanosystem where the transferred energy localizes at a given site. Because the electromagnetic wavelength is on a much larger microscale it is impossible to employ light-wave focusing for that purpose. In paper~\cite{10} the method to use phase modulation of an exciting femtosecond pulse is proposed. This method of localization exist due to the fact that polar excitation (surface plasmous) in inhomogeneous nanosystems tend to be localized with their oscillation frequency (and, consequently, phase) correlated with position~\cite{11, 12}.

The coherently controlled ultrafast energy localization in nanosystems introduced in paper~\cite{10}, can have applications in different fields that require directed nanosize-selective excitation.

In recent years the advances in scanning tunnelling microscopy (STM) made possible the manipulation of single atoms on top of a surface and the construction of quantum-nanometre scale structures of arbitrary shapes~\cite{13}. In particular, quantum corrals have been assembled by depositing a close line of atoms or molecules on Cu or noble metal surface~\cite{14}-\cite{17}. These surface have the property that for small wave vectors parallel to the surface a parabolic band of two-dimensional (2D) surface states uncoupled to bulk states exists~\cite{18}. In quantum corrals the STM tip can existed standing wave pattern of the one electron de Broglie waves.

In this paper we describe the thermal excitation of the de Broglie electron waves with attosecond laser pulses. With coherent control of the ultrashort laser pulses it is possible to concentrate the laser energy on the nanometer scale~\cite{10}. Following the results of paper~\cite{19} we will describe the temperature of the electron 2D gas with the help of the quantum hyperbolic heat transfer equation.
\section{The model}
In the following we consider the 2D heat transfer phenomena described by the equation~\cite{19}:
    \begin{equation}
    \frac{1}{v^2}\frac{\partial^2 T}{\partial t^2}+\frac{1}{D}\frac{\partial T}{\partial t}+\frac{2Vm}{\hbar^2}T=\nabla^2 T,\label{eq1}
    \end{equation}
    where $T$ is temperature of the 2D electron gas
    \begin{equation}
    T=T(x,y,t),\label{eq2}
    \end{equation}
    $D$ is the thermal diffusion coefficient, $V$ is the nonthermal potential and $m$ is the mass of the heat carriers--electrons.

We seek solution of Eq.~(\ref{eq1}) in the form
    \begin{equation}
    T(x,y,t)=e^{-\frac{t}{2\tau}}u(x,y,t).\label{eq3}
    \end{equation}
    After substitution of Eq.~(\ref{eq3}) to Eq.~(\ref{eq1}) one obtains
    \begin{equation}
    \frac{1}{v^2}\frac{\partial^2 u}{\partial t^2}-\nabla^2 u+qu=0,\label{eq4}
    \end{equation}
    where
    \begin{equation}
    q=\frac{2Vm}{\hbar^2}-\left(\frac{mv}{2\hbar}\right)^2\label{eq5}
    \end{equation}
    for $D=\frac{\hbar}{m}$.

    We can define the distortionless thermal wave as the wave which preserves the shape in the field of the potential $V$. The condition for conserving the shape can be formulated as
    \begin{equation}
    q=\frac{2Vm}{\hbar^2}-\left(\frac{mv}{2\hbar}\right)^2=0.\label{eq6}
    \end{equation}
    When Eq.~(\ref{eq6}) holds Eq.~(\ref{eq4}) has the form
    \begin{equation}
    \frac{1}{v^2}\frac{\partial^2 u}{\partial t^2}-\nabla^2 u=0\label{eq7}
    \end{equation}
    and condition~(\ref{eq6}) can be written as
    \begin{equation}
    V\tau\sim\hbar,\label{eq8}
    \end{equation}
    where $\tau$ is the relaxation time
    \begin{equation}
    \tau=\frac{\hbar}{mv^2}.\label{eq9}
    \end{equation}
We conclude that in the presence of the potential energy $V$ one can observe the undisturbed thermal wave only when the Heisenberg uncertainty relaxation~(\ref{eq8}) is fulfilled.

In the subsequent we will consider the thermal relaxation of the 2D electron gas contained in 2D circular quantum corral with the radius $r$. In that case in polar coordinates equation~(\ref{eq7}) has the form
    \begin{equation}
    \frac{1}{r}\frac{\partial}{\partial r}\left(r\frac{\partial u}{\partial r}\right) +\frac{1}{r^2}\frac{\partial^2 u}{\partial \theta^2}=\frac{1}{v^2}\frac{\partial^2 u}{\partial t^2},\label{eq10}
    \end{equation}
    where $0<r<a$, $-\pi<\theta<\pi$ with boundary condition
    \begin{eqnarray}
    u(r,\theta,0)&=&f(r,\theta), \quad 0<r<a, \quad-\pi<\theta\leq\pi,\nonumber \\
    \frac{\partial u}{\partial t}(r, \theta, 0)&=&g(r,\theta),\quad 0<r<a, \quad -\pi<\theta\leq\pi.\label{eq11}
    \end{eqnarray}
    Using separation of variables $u(r,t)=R(r)T(t)$ yields the solution
    \begin{eqnarray}
    u(r,\theta,t)&=&\sum_na_{on}J_0(\lambda_{on}r)\cos(\lambda_{on}vt)\nonumber\\
    &&\mbox{}+\sum_{m,n}a_{mn}J_m(\lambda_{mn}r)\cos(m\theta)\cos(\lambda_{mn}ct)\nonumber\\
    &&\mbox{}+\sum_{m,n}b_{mn}J_m(\lambda_{mn}r)\sin(m\theta)
    \cos(\lambda_{mn}ct)\nonumber\\
    &&\mbox{}+\sum_nA_{on}J_0(\lambda_{on}r)\sin(\lambda_{on}ct)\nonumber\\
    &&\mbox{}+\sum_{m,n}A_{mn}J_m(\lambda_{mn}r)\cos(m\theta)
    \sin(\lambda_{mn}ct)\nonumber\\
    &&\mbox{}+\sum_{m,n}B_{mn}J_m(\lambda_{mn}r)\sin(m\theta)
    \sin(\lambda_{mn}ct),\label{eq12}
    \end{eqnarray}
where $J_m$ represents the m-th Bessel function of the first kind, $\lambda_{mn}$ represents the n-th zero of $J_m$, and the coefficients $a_{on}, a_{mn},b_{mn}, A_{on}, A_{mn}$ and $B_{mn}$ can be find out in~\cite{20}.

In the subsequent we present the numerical solution of the Eq.~(\ref{eq7}) for the thermal wave with velocity $v=5\,10^{-3}c$, $c=$  light velocity. Considering formula~(\ref{eq9}) for relaxation time we obtain
    \begin{equation}
    \tau=\frac{\hbar}{mv^2}=160\,{\rm as}\label{eq13}
    \end{equation}
for $m=m_e\, (=0.51\,{\rm MeV})$ i.e. for electrons and for mean free path of the electrons in the 2D electron gas:
    \begin{equation}
    \lambda_{mfp}=v\tau\approx0.1\,{\rm nm}.\label{eq14}
    \end{equation}
From formula~(\ref{eq14}) we conclude that $\lambda_{mfp}$ is of the order of the de Broglie'a wave length of the electron. It means that for 2D electron gas in quantum stadium the hyperbolic quantum thermal equation can be applied~\cite{21}.

In Fig.~1 we present the shape of the laser pulse for circular
quantum stadium with $r=1$~nm and in Figs. 2, 3, 4 the numerical
solution of the Eq.~(\ref{eq10}) for the boundary
conditions~(\ref{eq11}) with $r=5, 10, 70$~nm respectively. As can
be easily seen in quantum stadium (corrals) the thermal wave is
propagated.

\section{Conclusions}
In this paper in the frame of the quantum hyperbolic heat transfer equation the study of the thermal transport in quantum stadium is investigated. It is shown that at the limit of quantum heat transport i.e. when the mean free path of the electrons is equal to its de Broglie'a wavelength in quantum corrals the thermal waves can be generated. As the result one can argue that the observed electron standing waves in quantum corrals are the precursors of the standing quantum thermal waves with the same shape and structures.

\begin{figure}[p]
\includegraphics[width=12cm]{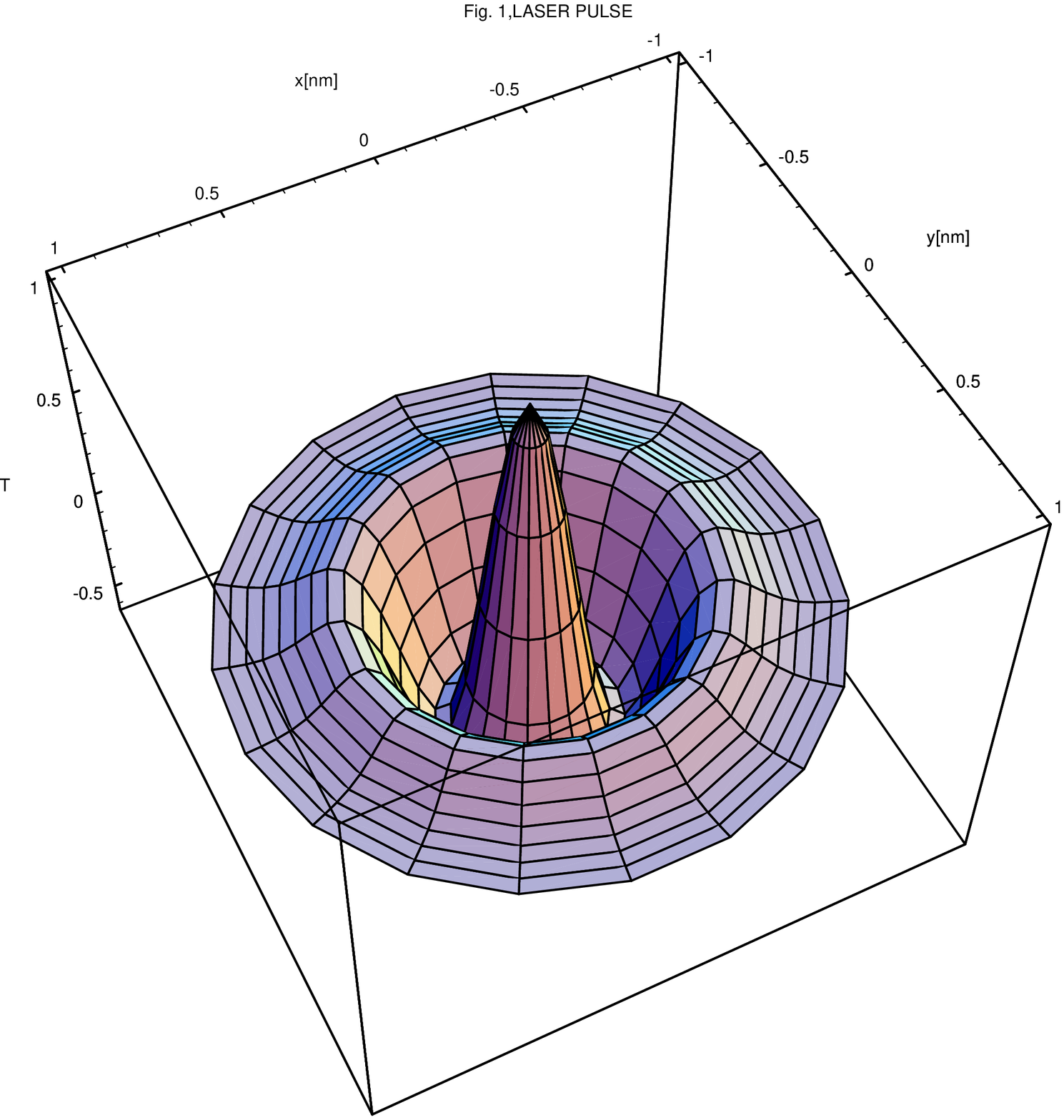}
\caption{The laser pulse}
\end{figure}

\begin{figure}[p]
\includegraphics[width=12cm]{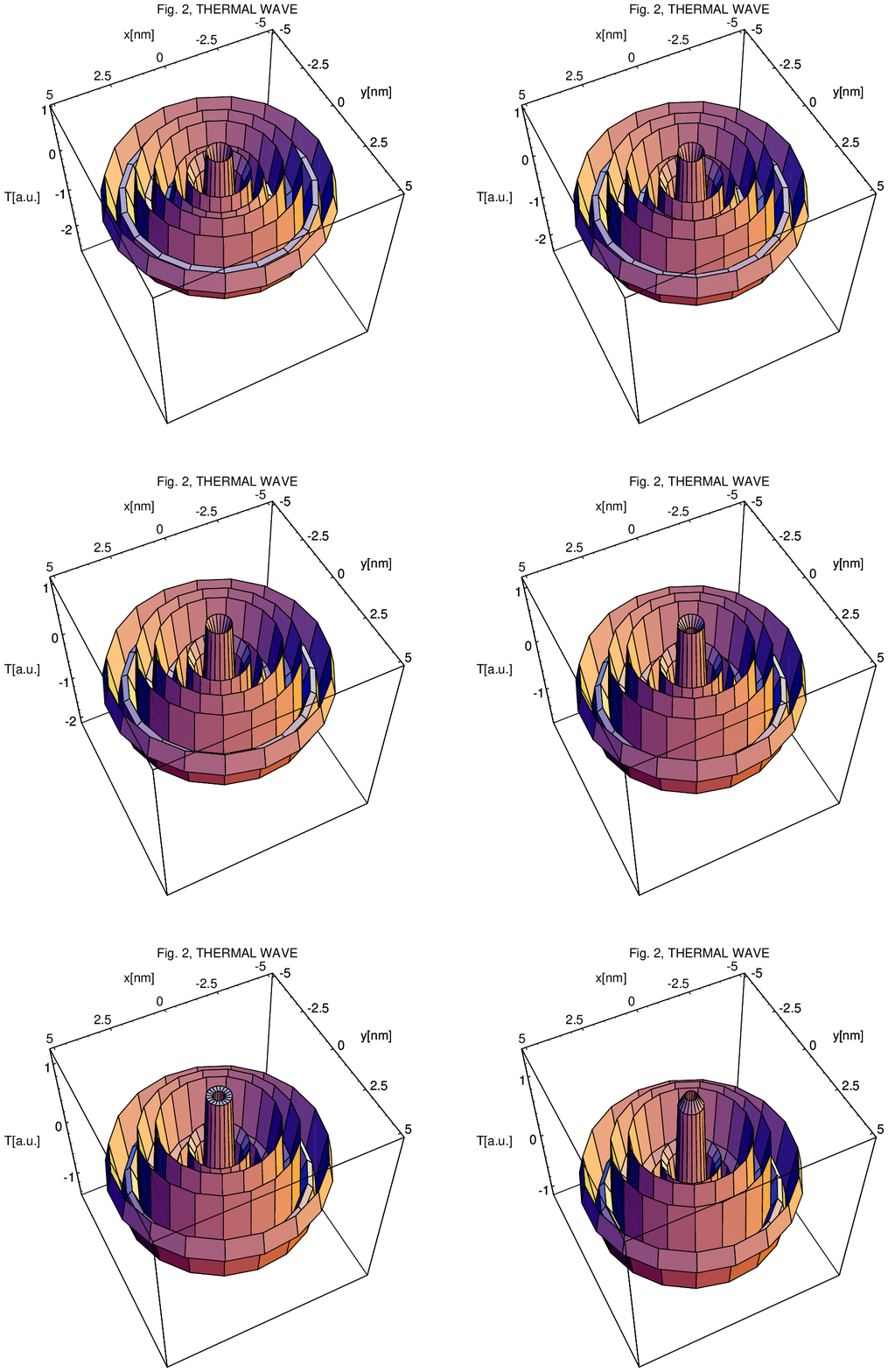}
\caption{Thermal wave in circular stadium with $r= 5$ nm}
\end{figure}
\begin{figure}[p]
\includegraphics[width=12cm]{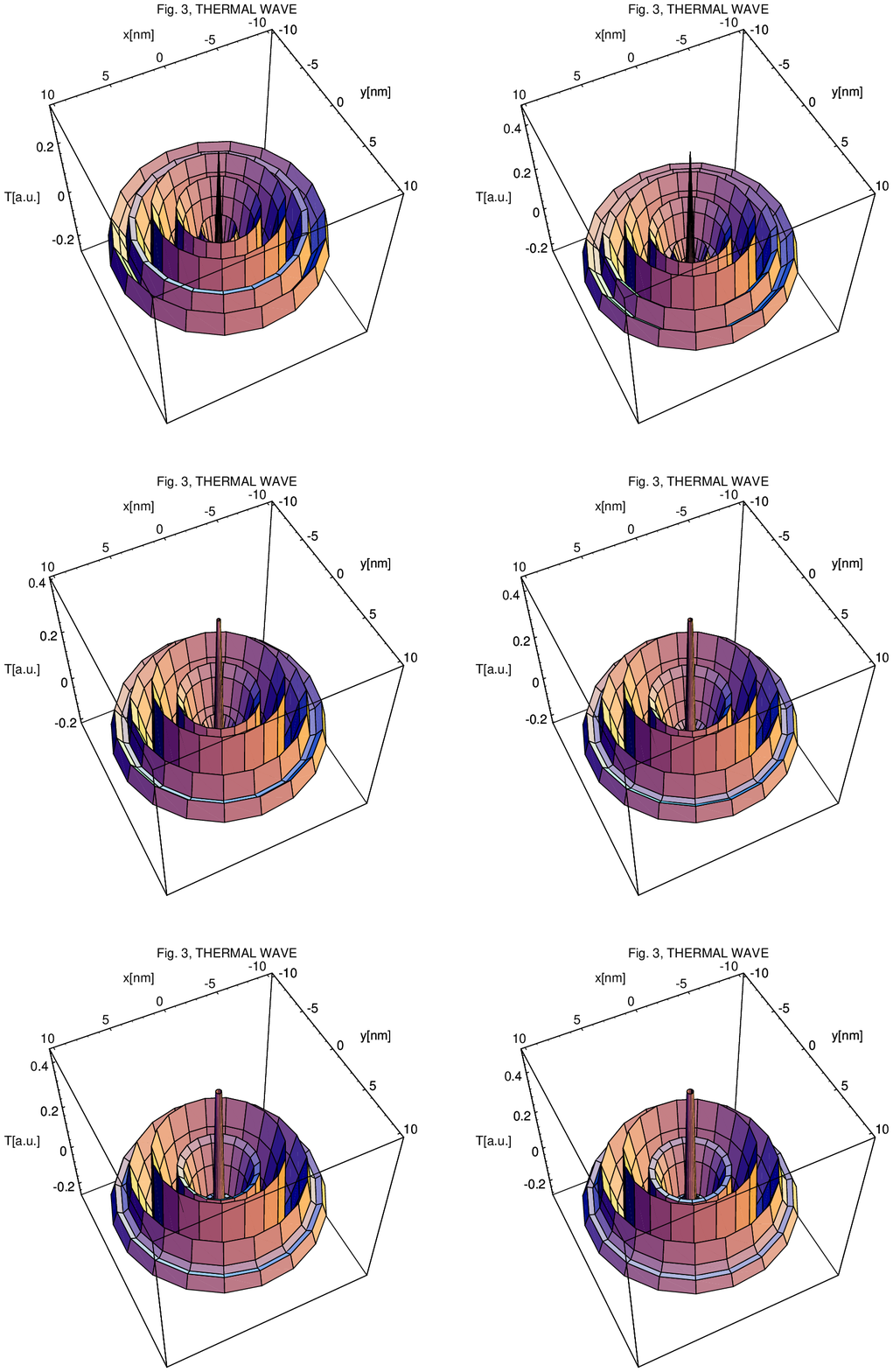}
\caption{Thermal wave in circular stadium with $r= 10$ nm}
\end{figure}
\begin{figure}[p]
\includegraphics[width=12cm]{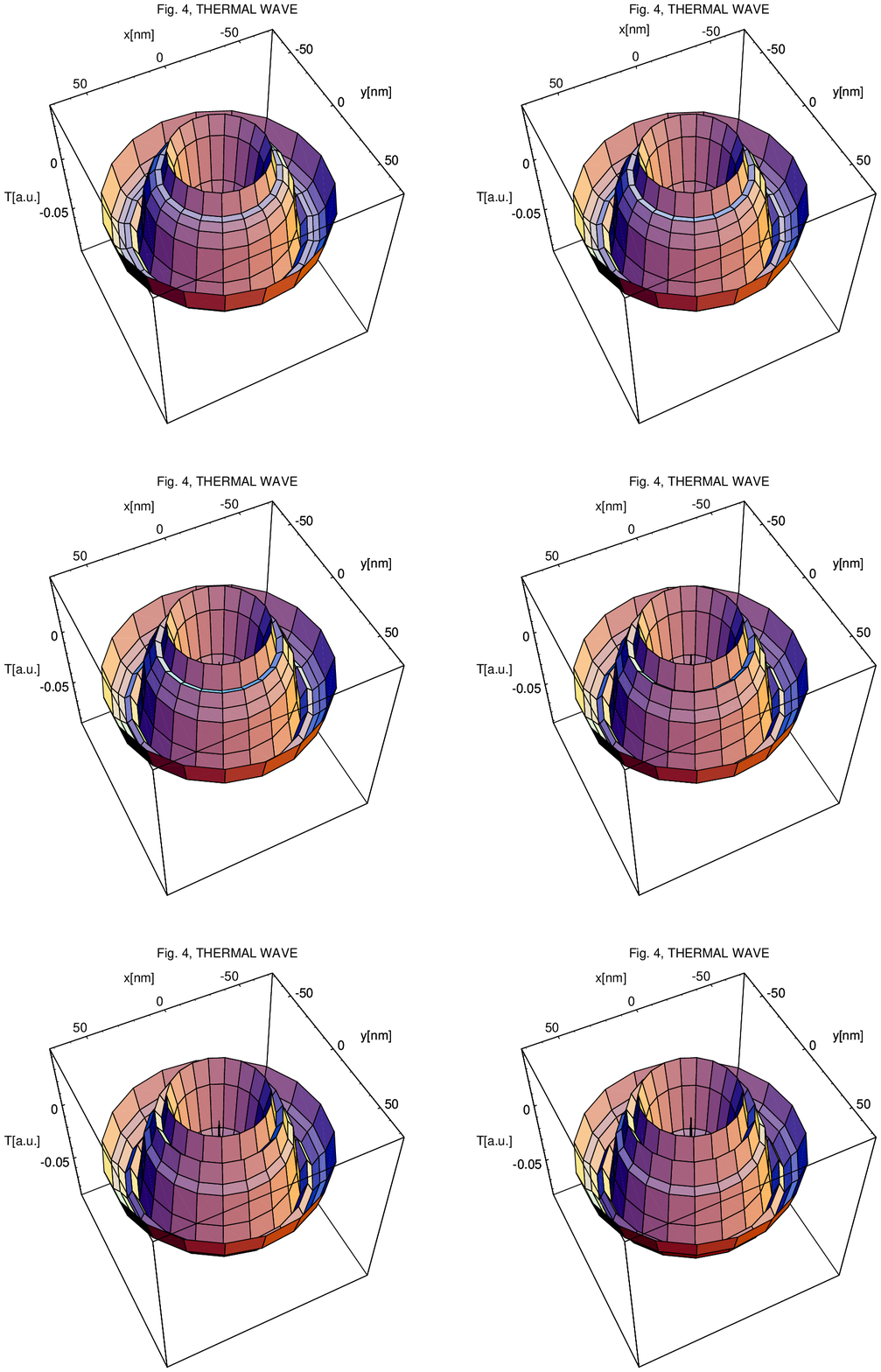}
\caption{Thermal wave in circular stadium with $r= 70$ nm}
\end{figure}
\end{document}